\documentclass[showkeys]{revtex4}
\usepackage[utf8]{inputenc}

\usepackage{amsmath}
\usepackage{graphicx}
\usepackage{color}
\usepackage{rotating}
\usepackage{array}
\usepackage{wrapfig}

\begin{document}

\title{
Scalar model of glueball in nonperturbative quantisation à la Heisenberg
}

\author{Vladimir Dzhunushaliev}
\email{v.dzhunushaliev@gmail.com}
\author{Arislan Makhmudov}
\email{arslan.biz@gmail.com}
\affiliation{IETP, Al-Farabi KazNU, Almaty, 050040, Kazakhstan}
\affiliation{Dept. Theor. Phys., KazNU, Almaty, 050040, Kazakhstan}

\begin{abstract}
A scalar model of glueball is considered. The model is based on two scalar fields approximation for SU(3) non-Abelian Lagrangian.  The approach to approximation makes use of the assumption that 2 and 4-points Green's functions are described in terms of some two scalar fields. The model is described via non-perturbative method due to value of coupling constant, which does not permit us using of Feynman diagrams and therefore of perturbative methods. Asymptotical behaviour of the scalar fields are obtained. Profiles of these fileds calculated for a range of values of a parameter of the problem is given. Detailed numerical investigation of corresponding equations describing this model is performed. The dependence of the glueball mass vs parameters of scalar fields is shown. Comparison of characteristics of glueball obtained in our two-scalar model and predictions of other models and experimental data for glueball is performed.
\end{abstract}

\date{\today}
\keywords{non-perturbative quantization, glueball, scalar field approximation}
\maketitle

\section{Introduction}
\label{sec-i}

Nowadays, one of the main problems of quantum chromodynamics (QCD) is the problem of glueball – a controversial object whose existence is fully admitted, though details of its model are still under question. Glueball represents a model for a new hadron which is created by self-interacting  gluonic fields. In a sense, a glueball  can be imagined as a blob of gluonic fields, and this causes difficulties in working with the model. Particularly, the model of glueball can be described only in the realm of QCD, since in this case we deal with gluonic fields. The non–dimensional coupling constant of SU(3) nonabelian gauge theory is the base for QCD, and the latter being strongly nonlinear theory requires the constant to be greater than 1. Thus, mathematically, this condition stipulates that we can not use Feynman diagram method, namely, the perturbative quantum field theory. The matter is that the method can be used only in case of weak interactions, in which dimensionless coupling constant $<1$. Therefore, the problem lies in that for description of the model we must use non-perturbative methods of quantum field theory. For example, in Ref's \cite{Brunner:2015oqa, Brunner:2015yha} the authors consider the glueball decay rates in the Witten-Sakai-Sugimoto model, a holographic top-down approach for QCD with chiral quarks in Witten’s holographic model of nonsupersymmetric Yang-Mills theory.

At the present time, experimental search of lightest glueball is conducted based on assessments of its mass, which lies in the range of 1000–1700 MeV (see Ref.~\cite{Ochs:2013gi} for the experimental and theoretical status of glueball). Although, it is necessary to mention that experimental search did not reveal any strong evidence of existence of glueball.

In Ref's~\cite{Dzhunushaliev:2015hoa, Dzhunushaliev:2003sq} an approximated approach was proposed for the non-perturbative quantisation in QCD. In this approach given SU(3) Lagrangian is turned into Lagrangian of two scalar fields using some assumptions. These fields describe two and four point Green functions entering the initial Lagrangian. Hence, we obtain simplified description of gluonic fields in the form of two scalar fields.

\section{Scalar toy model of glueball}

A glueball is a hypothetical composite particle that consists solely of non-Abelian SU(3) gauge field, without valence quarks. The existence of a glueball is consequence of the self-interaction of gluons within quantum chromodynamics. Nonlinear self-interaction of gluons in QCD leads to possibility of the existence of a color-neutral state made of gluons only, which was called glueball. Glueball is also thought as a bound state of gluons, and it's properties cannot be described within a perturbative approach to QCD. Glueball remains an obscure object over thirty years after QCD was used to predict such a state. It is well known that the gluon condensate, from which glueball is thought to be made of,  can only be determined in a nonperturbative formulation of QCD. So far various attempts have been made to determine gluon condensate from first principles \cite{Banks}, \cite{Shifman:1978bx}. We refer the reader to Ref.~\cite{Mathieu:2008me} for more details.

Our main idea is to write an effective Lagrangian. This Lagrangian is obtained from the Lagrangian of SU(3) non-Abelian gauge theory. In order to do this, we first separate SU(3) color degrees of freedom into two parts: subgroup~SU(2) $\subset$ SU(3) and coset SU(3)/SU(2). Then we average the SU(3) Lagrangian using some assumptions and approximation. Our approximation is based on the main assumption that the 2 and 4-points Green's functions are described in terms of some scalar fields $\phi$ and $\chi$ due to the following relations:
\begin{align}
	\left( G_2 \right)^{ab}_{\mu \nu}(x, x) & =
	\left\langle
		A^a_\mu (x) A^b_\nu (x)
	\right\rangle
	& \approx &
	m_1^2 - C^{ab}_{\mu \nu} \phi^2(x) ,
\label{qcd-gl-10}\\
	\left( F_2 \right)^{ab}_{\mu \nu}(x, x) & =
	\left\langle
		\partial_\mu A^a_\alpha (x) \partial_\nu A^b_\beta (x)
	\right\rangle
	& \approx &
	D^{ab}_{\alpha \beta} \partial_\mu \phi(x) \partial_\nu \phi(x) ,
\label{qcd-gl-10a}\\
	\left( G_4 \right)^{abcd}_{\mu \nu \rho \sigma}(x, x, x, x) &=
	\left\langle
		A^a_\mu (x) A^b_\nu (x) A^c_\rho (x) A^d_\sigma (x)
	\right\rangle
	&\approx &
	\left\langle
		A^a_\mu (x) A^b_\nu (x)
	\right\rangle
	\left\langle
		A^c_\rho (x) A^d_\sigma (x)
	\right\rangle
\label{qcd-gl-20}\\
	\left( G_2 \right)^{mn}_{\mu \nu}(x, x) &=
	\left\langle
		A^m_\mu (x) A^n_\nu (x)
	\right\rangle
	&\approx &
	C^{mn}_{\mu \nu} \chi^2(x) ,
\label{qcd-gl-30}\\
	\left( F_2 \right)^{mn}_{\mu \nu}(x, x) & =
	\left\langle
		\partial_\mu A^m_\alpha (x) \partial_\nu A^n_\beta (x)
	\right\rangle
	& \approx &
	D^{mn}_{\alpha \beta} \partial_\mu \phi(x) \partial_\nu \phi(x) ,
\label{qcd-gl-30a}\\
	\left( G_4 \right)^{mnpq}_{\mu \nu \rho \sigma}(x, x, x, x) &=
	\left\langle
		A^m_\mu (x) A^n_\nu (x) A^p_\rho (x) A^q_\sigma (x)
	\right\rangle
	&\approx & \;
	\left\langle
		A^m_\mu (x) A^n_\nu (x) - m_2^2
	\right\rangle
	\left\langle
		A^p_\rho (x) A^q_\sigma (x) - m_2^2
	\right\rangle - m_2^4,
\label{qcd-gl-40}\\
	\left( G_4 \right)^{ab mn}_{\mu \nu \rho \sigma}(x, x, x, x) &=
	\left\langle
		A^a_\mu (x) A^b_\nu (x) A^m_\rho (x) A^n_\sigma (x)
	\right\rangle
	&\approx & \;
	C^{ab mn}_{\mu \nu \rho \sigma} \phi^2(x) \chi^2(x)
\label{qcd-gl-41}
\end{align}
where $a,b,c,d = 1,2,3$ are SU(2) indices, $m,n,p,q = 4,5,6,7,8$ are coset indices, and
$C^{ab,mn}_{\mu\nu}, D^{ab,mn}_{\mu\nu},C^{ab mn}_{\mu \nu \rho \sigma}$ and $m_{1,2}$ are closure constants. Similarly, we see that  for turbulence modeling we have to introduce some closure constants. An effective Lagrangian then becomes
\begin{equation}
	\frac{g^2}{\hbar c} \mathcal L_{\rm{eff}} =
  \frac{g^2}{\hbar c} \left\langle \mathcal L_{SU(3)} \right\rangle =
	\frac{1}{2} \nabla_\mu \phi \nabla^\mu \phi -
	\frac{\lambda_1}{4} \left(
		 \phi ^2 - m_1^2
	\right)^2 +
	\frac{1}{2} \nabla_\mu \chi \nabla^\mu \chi -
	\frac{\lambda_2}{4} \left(
		\chi^2 - m_2^2
	\right)^2 + \frac{\lambda_2}{4} m_2^4 -
	\frac{1}{2}  \phi^2 \chi^2,
\label{qcd-gl-50}
\end{equation}
where $g$ is the dimesionless coupling constant; $\lambda_{1,2}$, $m_{1,2}$ are some parameters; the signature of the spacetime metrics is $(+,-,-,-)$. The effective Lagrangian \eqref{qcd-gl-50} is an approximation to the nonperturbatively quantized SU(3) gauge theory.

Quantities entering the Lagrangian \eqref{qcd-gl-50} have the following meanings and origins:
\begin{itemize}
	\item the scalar fields $\phi$ and $\chi$ describe nonperturbatively quantized SU(2) and coset SU(3)/SU(2) degrees of freedom, correspondingly;
	\item the terms $\nabla_\mu \phi \nabla^\mu \phi$ and
	$\nabla_\mu \chi \nabla^\mu \chi$ are the result of the nonperturbative quantum averaging of
	$( \nabla_\mu A^B_\nu )^2$ in the initial SU(3) Lagrangian;
	\item the terms $\phi^4$ and $\chi^4$ are the result of the nonperturbative quantum averaging of $f^{ABC} f^{AMN} A^B_\mu A^C_\nu A^{M \mu} A^{N \nu}$;
	\item the term $\phi^2 \chi^2$  is the result of the nonperturbative quantum averaging of $f^{Aab} f^{Amn} A^a_\mu A^b_\nu A^{m \mu} A^{n \nu}$;
  \item $m_{1,2}$ appear to be closure coefficients;
	\item the terms $\phi^2 m_1^2$,	$\chi^2 m_2^2$ arise due to the closure coefficients $m_{1,2}$.
\end{itemize}
Using the Lagrangian \eqref{qcd-gl-50}, we derive the associated field equations describing glueball in the following form:
\begin{eqnarray}
  \partial_\mu \partial^\mu \phi &=&
  - \phi \left[ \chi^2 + \lambda_1
  \left(
    \phi^2 - m_1^2
  \right) \right],
\label{qcd-gl-60}\\
  \partial_\mu \partial^\mu \chi &=&
  - \chi \left[ \phi^2 + \lambda_2
  \left(
    \chi^2 - m_2^2
  \right) \right].
\label{qcd-gl-70}
\end{eqnarray}
In Ref. \cite{Dzhunushaliev:2003sq} the spherically symmetric solution to equations \eqref{qcd-gl-60} and \eqref{qcd-gl-70} is considered. This ball serves as a nonperturbative scalar model of a glueball. It is  shown in Ref.~\cite{Dzhunushaliev:2003sq}  that such solutions with finite energy do exist.

Here we want to investigate these solutions in more details.

\section{Two scalar model of glueball}
\label{sec-mo}

Our model is described by a system of equations for coupled scalar fields \eqref{qcd-gl-60}-\eqref{qcd-gl-70}. In order to describe glueball we consider the spherical symmetric case. Let us rewrite the equations in the dimensionless form
\begin{eqnarray}
  {\phi}''+\frac{2}{x}{\phi}' &=& \phi[\chi^2 +
  \lambda_1(\phi^2-m_1^2)],
\label{1-25}\\
  {\chi}''+\frac{2}{x}{\chi}' &=& \chi[\phi^2 +
  \lambda_2(\chi^2 - m_2^2)],
\label{1-28}
\end{eqnarray}
here redefinitions were made
$
\phi \to \phi/\phi(0)$, $\chi \to \chi/\phi(0), m_{1,2} \to m_{1,2}/ \phi(0)
$, and the dimensionless coordinate $x = r \phi(0)$ is introduced. The field equations \eqref{1-25}-\eqref{1-28} can not be solved analytically and therefore we will solve them using numerical methods.

\subsection{Behaviour of the scalar fields at the origin and at infinity}
\label{sec-asb}

Numerical analysis shows that the asymptotical behaviour of the scalar fields is as follows
\begin{equation}
 \phi(x) \rightarrow m_1,
 \chi(x) \rightarrow 0.
 \label{1-28a}
\end{equation}
Taking into account the asymptotical behaviour \eqref{1-28a} and equations \eqref{1-25} and \eqref{1-28} we can find the asymptotical behaviour of the scalar fields
\begin{eqnarray}
  \phi(x)&\approx&m_1-
  \phi_{\infty} \frac{e^{- x \sqrt{2 \lambda_1 m_1^2}}}{x} ,
\label{1-29}\\
  \chi(x)&\approx& \chi_\infty \frac{e^{ - x \sqrt{m_1^2-\lambda_2 m_2^2}}}{x},
\label{1-30}
\end{eqnarray}
where $\phi_{\infty}$, $\chi_{\infty}$ are some constants.
Near the origin, the scalar fields change according to the following laws:
\begin{eqnarray}
  \phi(x)&=&\phi_0 + \frac{\phi_2 x^2}{2} + \ldots ,
\label{1-31}\\
  \chi(x)&=&\chi_0 + \frac{\chi_2 x^2}{2} + \ldots
\label{1-32}
\end{eqnarray}
where
\begin{eqnarray}
  \phi_2 &=& \phi_0 \left[
   \chi_0^2 + \lambda_1 \left(
    \phi_0^2 - m_1^2
   \right)
  \right] ,
\label{1-31a}\\
  \chi_2 &=& \chi_0 \left[
    \chi_0^2 + \lambda_2 \left(
      \chi_0^2 - m_2^2
    \right)
  \right].
\label{1-32a}
\end{eqnarray}

\subsection{Numerical solution}
\label{sec-sol}

The numerical analysis shows us that regular solutions of \eqref{1-25}-\eqref{1-28} exist for some special case of parameters $m_{1,2}$ only. It means that we have to consider these equations' set as a non-linear eigenvalue problem for eigenvalues $m_{1,2}$ and eigenfunctions $\phi(x), \chi(x)$. The solution to the system of equations \eqref{1-25}-\eqref{1-28} was obtained via "step by step" method, when approximate solution is improved in every following step.

\begin{figure}[ht]
\begin{minipage}[ht]{.45\linewidth}
\vspace{1cm}
\begin{center}
	\fbox{
	\includegraphics[width=.9\linewidth]{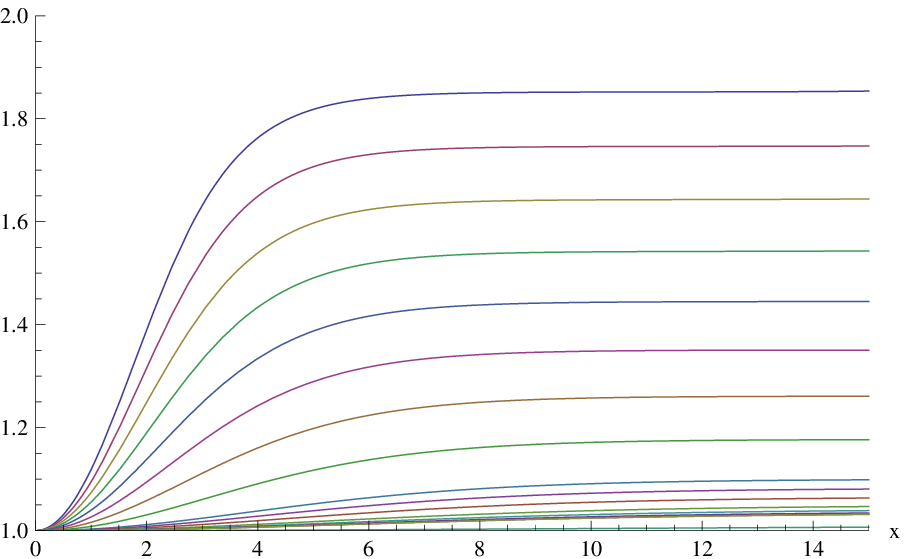}
	}
\end{center}
\caption{Bunch of  $\phi$(x) curves calculated for a range of values of $\chi_0$ from 0.05 to 1. The bottommost curve  corresponds to the value of $\chi_0=0.05$, every curve higher than this corresponds to greater values of $\chi_0$ ending up with the topmost curve for $\chi_0=1$. }
\label{bunch-of-phi-of-x.eps}
\end{minipage}\hfill
\begin{minipage}[ht]{.45\linewidth}
\begin{center}
	\fbox{
	\includegraphics[width=.9\linewidth]{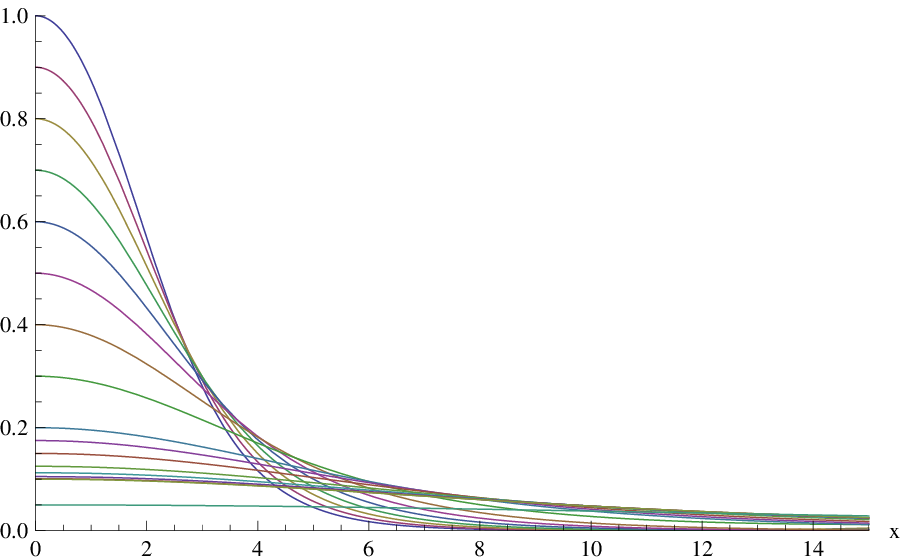}
	}
\end{center}
\caption{Bunch of $\chi(x)$ curves calculated for the range of values of $\chi_0$ from 0.05 to 1. It can be seen  from initial value of a curve that the topmost curve has been calculated for $\chi_0$=1, the bottommost curve corresponds to $\chi_0$=0.05.}
\label{bunch-of-chi-of-x.eps}
\end{minipage}
\end{figure}

Firstly, we expanded the system of equations near the origin using Taylor series, and got the boundary conditions
\begin{equation}
 \phi(0) = 1,\; \phi'(0)=0,
 \chi(0) =\chi_0,\; \chi'(0)=0.
\label{1-35}
\end{equation}
The system of equations \eqref{1-25}-\eqref{1-28}  was solved as a non-linear eigenvalue problem in which values of the following parameters were sought: $m_{1,2}$. Other parameters except $\chi_0$ in the system of equations were set to appropriate values:
\begin{equation}
	\lambda_1=0.1, \; \lambda_2=1.0,
\label{1-40}
\end{equation}
while $\chi_0$ was varied in the range of values from $0.05$ to $50$. For every $\chi_0$ we got appropriate eigenvalues of $m_{1,2}$ which are given in Table \ref{obtained-data}. Profiles of solutions for a range of $\chi_0$ values are given in Fig.~\ref{bunch-of-phi-of-x.eps}-\ref{bunch-of-chi-of-x.eps}.

\begin{table}[ht]
\centering
\begin{minipage}{0.4\linewidth}
  \centering
\begin{tabular}{ | m{2cm} | m{2cm} | m{2cm} | }
  \hline
  $\chi_0$&$m_1$&$m_2$\\
  \hline
  0.05&1.01077&1.01019\\
  \hline
  0.1&1.03487&1.03189\\
  \hline
  0.2&1.1&1.08769\\
  \hline
  0.3&1.17779&1.1501\\
  \hline
  0.4&1.2612&1.21803\\
  \hline
  0.5&1.3509&1.28827\\
  \hline
  0.6&1.44491&1.36096\\
  \hline
  0.7&1.54239&1.43576\\
  \hline
  0.8&1.643&1.51235\\
  \hline
  0.9&1.74629&1.59053\\
  \hline
  1&1.85201&1.67014\\
  \hline
  2&3.00371&2.52084\\
  \hline
  5&6.9365&5.32339\\
  \hline
  10&14.0095&10.2078\\
  \hline
  15&21.307&15.1572\\
  \hline
  25&36.1893&25.1091\\
  \hline
  30&43.7127&30.0954\\
  \hline
  50&74.1089&50.0653\\
  \hline
\end{tabular}
\caption{The eigenvalues $m_{1,2}$ calculated from equations \eqref{1-25}-\eqref{1-28} for corresponding values of $\chi_0$.}
\label{obtained-data}
\end{minipage} \hfill
\begin{minipage}{0.4\linewidth}
\centering
\begin{tabular}{ | m{2cm} | m{2cm} | m{2cm} | }
\hline
 $\chi_0$&$x_{\phi,0}$&$x_{\chi,0}$\\
  \hline
  0.05&12.1963&17.1703\\
  \hline
  0.1&7.39187&9.61754\\
  \hline
  0.2&4.85681&5.78707\\
  \hline
  0.3&3.92591&4.46546\\
  \hline
  0.4&3.3682&3.7237\\
  \hline
  0.5&3.01152&3.2655\\
  \hline
  0.6&2.75011&2.94117\\
  \hline
  0.7&2.54635&2.69574\\
  \hline
  0.8&2.38164&2.50203\\
  \hline
  0.9&2.24421&2.34373\\
  \hline
  1&2.1271&2.21116\\
  \hline
  2&1.4691&1.50183\\
  \hline
  3&1.15977&1.18304\\
  \hline
  5&0.839859&0.856706\\
  \hline
  10&0.520063&0.530226\\
  \hline
  15&0.385755&0.392831\\
  \hline
  25&0.260804&0.265031\\
  \hline
  30&0.22602&0.229494\\
  \hline
  50&0.150125&0.152079\\
  \hline
\end{tabular}
\caption{Characterstic sizes $x_{\phi,0}$ , $x_{\chi,0}$ of regions of $\phi$(x) and $\chi$(x) fields correspondingly calculated for different values of $\chi_0$.}
\label{tbl-char-sz-chi-phi}

\end{minipage}
\hfill
\end{table}

\begin{figure}[ht]
\begin{minipage}{.45\linewidth}
	\fbox{	
  \includegraphics[width=.9\linewidth]{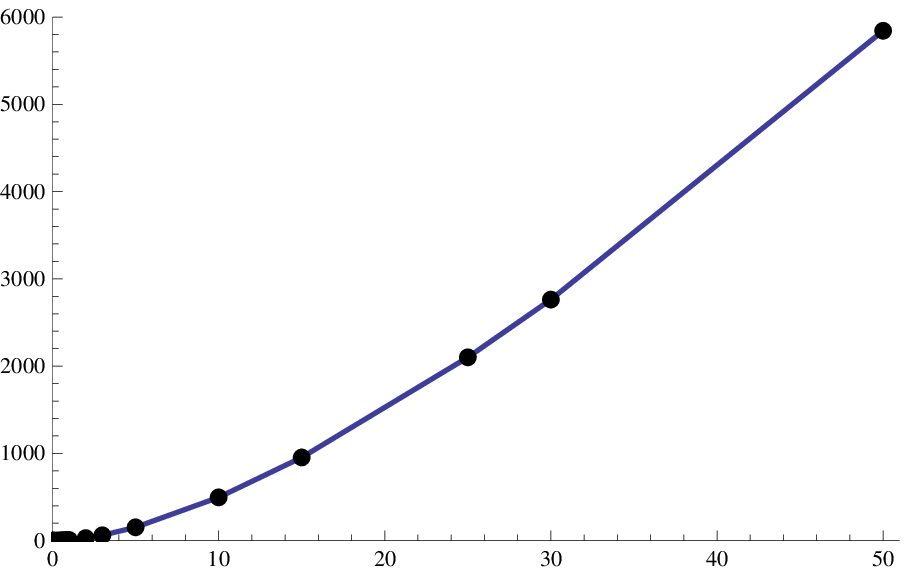}
	}
\caption{The dimensional glueball energy  calculated for values $\chi_0$,
$m_{1,2}$ given in Table \ref{obtained-data}.}
\label{energy-of-chi-0}
\end{minipage}\hfill
\begin{minipage}{.45\linewidth}
\vspace{.9cm}
	\fbox{	
  \includegraphics[width=.9\linewidth]{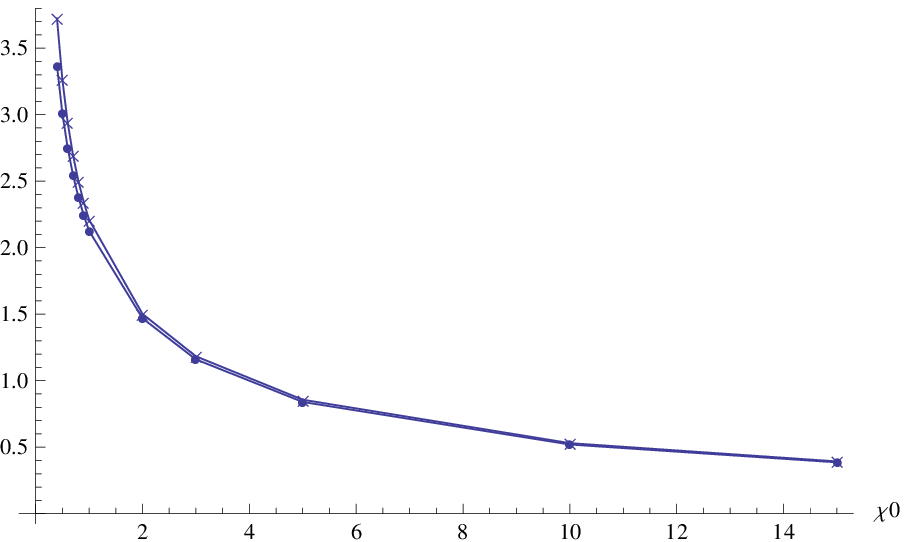}
	}
\caption{Characteristic sizes of regions of $\phi(x)$, $\chi(x)$ fields calculated for different values of $\chi_0$ given in Table \ref{tbl-char-sz-chi-phi}. Dotted curve corresponds to the characteristic sizes of $\phi(x)$, while crossed one corresponds to the sizes of $\chi(x)$ field.}
\label{chrc-sz-chi-phi}
\end{minipage}
\end{figure}

The dimensionless energy density  of the presented solution is
\begin{equation}
	\frac{g^2}{\hbar c} \frac{1}{\phi_0^4} \varepsilon(r) =
  \tilde \varepsilon(x) =
	\frac{1}{2} {\phi'}^2(x) + \frac{1}{2} {f'}^2(x) + \frac{\lambda_1}{4}
  \left( \phi^2(x) - m^2_1 \right)^2 +
  \frac{\lambda_2}{4} f^2(x) \left( f^2(x) - 2 m^2_2 \right) +
	\frac{1}{2} f^2(x) \phi^2(x)	  	
\label{sec5:25}
\end{equation}
Thus the dimensionless glueball energy is
\begin{equation}
	\frac{g^2}{\hbar c} \frac{1}{\phi_0} W =
  \tilde W = 4 \pi \int\limits^\infty_0 x^2
	\left[
    \frac{1}{2} {f'}^2 + \frac{1}{2} {\phi'}^2 +
    \frac{\lambda_1}{4} \left( \phi^2 - m^2_1 \right)^2 +
	  \frac{\lambda_2}{4} f^2 \left( f^2 - 2m^2_2 \right) +
	  \frac{1}{2} f^2 \phi^2
	\right] dx
\label{sec5-30}
\end{equation}
The glueball energies were calculated for each set of $\chi_0$ together with eigenvalues given in Table \ref{obtained-data}, and plotted to
Fig.~\ref{energy-of-chi-0}.

Also, we have calculated the characteristic sizes $x_{\phi, 0}, x_{\chi, 0}$ (they characterize the glueball size) for every value $\chi_0$ in the following way
\begin{eqnarray}
	\phi(x_{\phi, 0}) &=& \phi(0) + \frac{m_1 - \phi(0)}{2},
\label{sec5-40}\\
	\chi(x_{\chi, 0}) &=& \frac{\chi(0)}{2}.
\label{sec5-50}
\end{eqnarray}
These values are presented in Table \ref{tbl-char-sz-chi-phi}, and plotted to Fig.~\ref{chrc-sz-chi-phi}.

\section{Comparison of numerical values of our model with that of other models}
\label{sec-comp}

Now we want to compare the energy of glueball of our model with that of obtained in other models. We considered the models reviewed in Ref. \cite{Ochs:2013gi} which gives the mass for the lightest glueball within the range of 1000 - 1700 MeV. For comparison, we considered one of them with the mass of 1500 MeV. First, we calculated $\phi_0$ parameter from the energy taken from Ref \cite{Ochs:2013gi}. After that, we calculated the same constant from characteristic size of a glueball which supposedly is the radius of a proton, i.e. 1 fermi, Eq. \eqref{1-45}.
\begin{eqnarray}
 \phi(0)&=&\frac{g^2 m_g}{\hbar \; c \; \tilde{E}},
\label{1-40a}\\
 \phi(0) &=&\frac{x_0}{r_0},
\label{1-45}
\end{eqnarray}
where $m_g$=1500 MeV is the glueball mass; $g \approx 1$; $x_0$ is dimensionless characteristic radius of a glueball taken from our model, which is roughly taken equal to 10, we have taken this value because this number represents the characteristic size of the region where scalar fields $\phi$(x), $\chi$(x) are concentrated (i.e. radius of glueball), see Table \ref{tbl-char-sz-chi-phi}, Fig. \ref{chrc-sz-chi-phi} for details; $\tilde{E}$ is dimensionless energy calculated for given eigenvalues $m_{1,2}$ and parameter $\chi_0$; $r_0$ is the numerical value of radius of proton, namely, 1 fermi.

The value of $\phi_0$ we obtained for the mass of 1500 MeV and for $\chi_0$=0.05, $\phi_0\;\approx 10^{15\;}cm^{-1}$ and $\phi_0$ calculated from the data of our model is $\phi_0\;\approx 10^{16\;}cm^{-1}$. We can see that these values are in good agreement, taking into account that we compared the values qualitatively.

\section{Conclusions and discussion}
\label{sec-con}

Thus, we have investigated the solutions describing glueball approximately in the approach of two scalar fields. We have obtained the set of regular solutions having finite energy. Every solution has a good asymptotic behaviour, bringing us to the finite energy of this configuration of $\phi(x), \chi(x)$ fields. We have shown that when $\chi_0$ parameter is decreased, the region in which $\phi(x), \chi(x)$ fields are concentrated grows and correspondingly when the parameter is increased the sizes of the region decrease.
Physically, this means that in our approximate model the size of glueball depends on $\chi_0$ parameter in the way we described.  We have also investigated the dependence of glueball energy on $\chi_0$ parameter. We have shown that when $\chi_0 \to 0$ the energy decreases and correspondingly when $\chi_0 \to \infty$ the energy blows up.

It is significant to note that having dependence on $\chi_0$ parameter of the energy of glueball, one can investigate  statistical properties of glueball which has thermal contact with  thermostat.
In such case, glueball will play a role of statistical object , in which fluctuations of energy occur due to thermal contact with thermostat. Specificity of such investigation is in that we have statistical quantum object consisting not of quantum particles but of fluctuating quantum fields. From mathematical point of view this means that we have to calculate statistical sum for non-perturbative quantised object. Such problem represents a complicated issue as well as any other problem in the area of non-perturbative quantisation.

\section*{Acknowledgements}
We acknowledge support from a grant No.~3101/GF4 and No.~1626/GF3 in fundamental research in natural sciences by the Ministry of Education and Science of Kazakhstan.


\begin{thebibliography}{99}

\bibitem{Brunner:2015oqa}
  F.~Brünner, D.~Parganlija and A.~Rebhan,
  ``Glueball Decay Rates in the Witten-Sakai-Sugimoto Model,''
  Phys.\ Rev.\ D {\bf 91}, no. 10, 106002 (2015)
  [arXiv:1501.07906 [hep-ph]].

\bibitem{Brunner:2015yha}
  F.~Brünner and A.~Rebhan,
  ``Nonchiral enhancement of scalar glueball decay in the Witten-Sakai-Sugimoto model,''
  arXiv:1504.05815 [hep-ph].

\bibitem{Ochs:2013gi}
  W.~Ochs,
 ``The Status of Glueballs,''
  J.\ Phys.\ G {\bf 40}, 043001 (2013)
  [arXiv:1301.5183 [hep-ph]].

\bibitem{Dzhunushaliev:2015hoa}
  V.~Dzhunushaliev,
  ``Nonperturbative quantization: ideas, perspectives, and applications,''
  arXiv:1505.02747 [physics.gen-ph].

\bibitem{Dzhunushaliev:2003sq}
V.~Dzhunushaliev,
``Scalar model of the glueball,''
Hadronic J.\ Suppl.\  {\bf 19}, 185 (2004);
hep-ph/0312289.

\bibitem{Banks}
T. Banks, R. Horsley, H.R. Rubinstein, and U. Wolff,
Nucl. Phys. \textbf{B190}, 692 (1981);
A.Di Giacomo and G.C. Rossi,
Phys. Lett. B \textbf{100}, 481 (1981);
P.E. Rakow,
PoS LAT2005, 284 (2006) and references therein.

\bibitem{Shifman:1978bx}
  M.~A.~Shifman, A.~I.~Vainshtein, V.~I.~Zakharov,
  Nucl.\ Phys.\  {\bf B147}, 385-447 (1979).

\bibitem{Mathieu:2008me}
  V.~Mathieu, N.~Kochelev, V.~Vento,
  Int.\ J.\ Mod.\ Phys.\  {\bf E18}, 1-49 (2009).
  [arXiv:0810.4453 [hep-ph]].

\end{thebibliography}
\end{document}